\documentclass[11pt]{article}
\pdfoutput=1

\usepackage{graphicx}
\usepackage{amsmath}
\usepackage{amsfonts}
\usepackage{enumitem}
\usepackage{fancyvrb}
\usepackage[a4paper,top=3cm]{geometry}
\usepackage{color}
\usepackage{titlesec}
\usepackage{booktabs}
\usepackage{float}
\usepackage{todonotes}
\usepackage{setspace}
\usepackage{fixltx2e}
\usepackage{siunitx}
\usepackage[labelfont=bf]{caption}
\usepackage{epstopdf}
\usepackage{natbib}
\usepackage{natbibspacing}
\usepackage{etoolbox}
\usepackage[utf8]{inputenc}
\usepackage{datetime}
\usepackage{multirow}

\usepackage{lmodern}
\usepackage[T1]{fontenc}
\usepackage{microtype}

\usepackage[pdftex,bookmarks]{hyperref}

\usepackage{boxedminipage}
\hypersetup{colorlinks=true, linkcolor=blue, citecolor=blue, urlcolor=blue}

\setlist{itemsep=0pt,leftmargin=*}

\titlespacing{\section}{0pt}{0pt}{3pt}
\titlespacing{\subsection}{0pt}{0pt}{0pt}

\titlespacing{\section}{0pt}{*0}{*0}

\makeatletter

\pretocmd{\NAT@citex}{%
  \let\NAT@hyper@\NAT@hyper@citex
  \def\NAT@postnote{#2}%
  \setcounter{NAT@total@cites}{0}%
  \setcounter{NAT@count@cites}{0}%
  \forcsvlist{\stepcounter{NAT@total@cites}\@gobble}{#3}}{}{}
\newcounter{NAT@total@cites}
\newcounter{NAT@count@cites}
\def\NAT@postnote{}

\def\NAT@hyper@citex#1{%
  \stepcounter{NAT@count@cites}%
  \hyper@natlinkstart{\@citeb\@extra@b@citeb}#1%
  \ifnumequal{\value{NAT@count@cites}}{\value{NAT@total@cites}}
    {\ifNAT@swa\else\if*\NAT@postnote*\else%
     \NAT@cmt\NAT@postnote\global\def\NAT@postnote{}\fi\fi}{}%
  \ifNAT@swa\else\if\relax\NAT@date\relax
  \else\NAT@@close\global\let\NAT@nm\@empty\fi\fi% avoid compact citations
  \hyper@natlinkend}
\renewcommand\hyper@natlinkbreak[2]{#1}

\patchcmd{\NAT@citex}
  {\ifNAT@swa\else\if*#2*\else\NAT@cmt#2\fi
   \if\relax\NAT@date\relax\else\NAT@@close\fi\fi}{}{}{}
\patchcmd{\NAT@citex}
  {\if\relax\NAT@date\relax\NAT@def@citea\else\NAT@def@citea@close\fi}
  {\if\relax\NAT@date\relax\NAT@def@citea\else\NAT@def@citea@space\fi}{}{}

\makeatother

\begin{document}
\begin{center}
	{\LARGE\textbf{The cardiac Na\textsuperscript{+}/K\textsuperscript{+} ATPase: An updated, thermodynamically consistent model}} \\
	Michael Pan$^{1}$, Peter J. Gawthrop$^{1}$, Joseph Cursons$^{2}$, Kenneth Tran$^{3}$,  \mbox{and Edmund J. Crampin$^{1,4,5}$} \\[0cm]
\end{center}
{\small
	$^1$Systems Biology Laboratory, School of Mathematics and Statistics, and Department of Biomedical Engineering, Melbourne School of Engineering, University of Melbourne, Parkville, Victoria 3010, Australia. \\
	$^2$Bioinformatics Division, Walter and Eliza Hall Institute of Medical Research, Parkville, Victoria 3052, Australia. \\
	$^3$Auckland Bioengineering Institute, University of Auckland, Auckland, New Zealand \\
	$^4$ARC Centre of Excellence in Convergent Bio-Nano Science and Technology, Melbourne School of Engineering, University of Melbourne, Parkville, Victoria 3010, Australia. \\
	$^5$School of Medicine, University of Melbourne, Parkville, Victoria 3010, Australia }

\textbf{Abstract} \\
The Na$^+$/K$^+$ ATPase is an essential component of cardiac electrophysiology, maintaining physiological Na$^+$ and K$^+$ concentrations over successive heart beats. \citet{terkildsen_balance_2007} developed a model of the ventricular myocyte Na$^+$/K$^+$ ATPase to study extracellular potassium accumulation during ischaemia, demonstrating the ability to recapitulate a wide range of experimental data, but unfortunately there was no archived code associated with the original manuscript. Here we detail an updated version of the model and provide CellML and MATLAB code to ensure reproducibility and reusability. We note some errors within the original formulation which have been corrected to ensure that the model is thermodynamically consistent, and although this required some reparameterisation, the resulting model still provides a good fit to experimental measurements that demonstrate the dependence of Na$^+$/K$^+$ ATPase pumping rate upon membrane voltage and metabolite concentrations. To demonstrate thermodynamic consistency we also developed a bond graph version of the model. We hope that these models will be useful for community efforts to assemble a whole-cell cardiomyocyte model which facilitates the investigation of cellular energetics.

\section{Introduction}
Cardiomyocytes maintain Na$^+$ and K$^+$ ions within their physiological concentration range, in part by using Na$^+$/K$^+$ ATPases located on their plasma membranes. The Na$^+$/K$^+$ ATPase is an electrogenic ion pump that uses energy from ATP hydrolysis to drive the transport of Na$^+$ and K$^+$ ions against an electrochemical gradient. A previous model of the cardiomyocyte Na$^+$/K$^+$ ATPase was developed by \citet{terkildsen_balance_2007} and subsequently incorporated into a whole-cell cardiomyocyte model to demonstrate that reduced Na$^+$/K$^+$ ATPase activity plays a dominant role in extracellular potassium accumulation during ischaemia \citep{terkildsen_balance_2007,terkildsen_modelling_2006}. In this manuscript the Na$^+$/K$^+$ ATPase model of \citet{terkildsen_balance_2007} will subsequently be referred to as the Terkildsen \textit{et al.} model.

The Na$^+$/K$^+$ ATPase model presented in \citet{terkildsen_balance_2007} was based upon an earlier implementation which proposed thermodynamic constraints and a lumping scheme for model simplification \citep{smith_development_2004}. A key development was that the Terkildsen \textit{et al.} model could reproduce a wider range of data which captured the dependence of the pump current upon membrane voltage \citep{nakao_[na]_1989}, extracellular sodium \citep{nakao_[na]_1989}, intracellular sodium \citep{hansen_dependence_2002}, extracellular potassium \citep{nakao_[na]_1989} and MgATP \citep{friedrich_na+k+-atpase_1996}. Unfortunately the cycling velocity figures presented within the original paper are not reproducible using information supplied in the figure legends \citep[Fig. 2]{terkildsen_balance_2007} and code used to generate those figures was not publicly archived. These issues are exacerbated by apparent errors within the reported equations and parameter values (further described in \autoref{sec:modifications}) which result in physical and thermodynamic inconsistencies.

Here we address these issues, updating the model to ensure that it is thermodynamically consistent, and archiving MATLAB and CellML \citep{lloyd_cellml:_2004} code for reproducibility. This required the modification of several equations (\autoref{sec:modifications}) and re-parameterisation through fitting to the original data (\autoref{sec:raparameterisation}). To verify the physical plausibility of the updated model we have also developed a bond graph version \citep{oster_network_1971,gawthrop_energy-based_2014}, and we refer readers to \citet{gawthrop_metamodelling:_1996}; \citet{borutzky_bond_2010}; and \citet{gawthrop_bond-graph_2007} for further information on bond graph theory. Given the thermodynamic consistency of our updated model we believe that it is particularly well-suited for incorporation into community efforts for developing a thermodynamic model of a cardiomyocyte to ultimately study whole-heart cardiac energetics. 

\section{Modifications}
\label{sec:modifications}
\begin{figure}
\centering
\includegraphics[width=\linewidth]{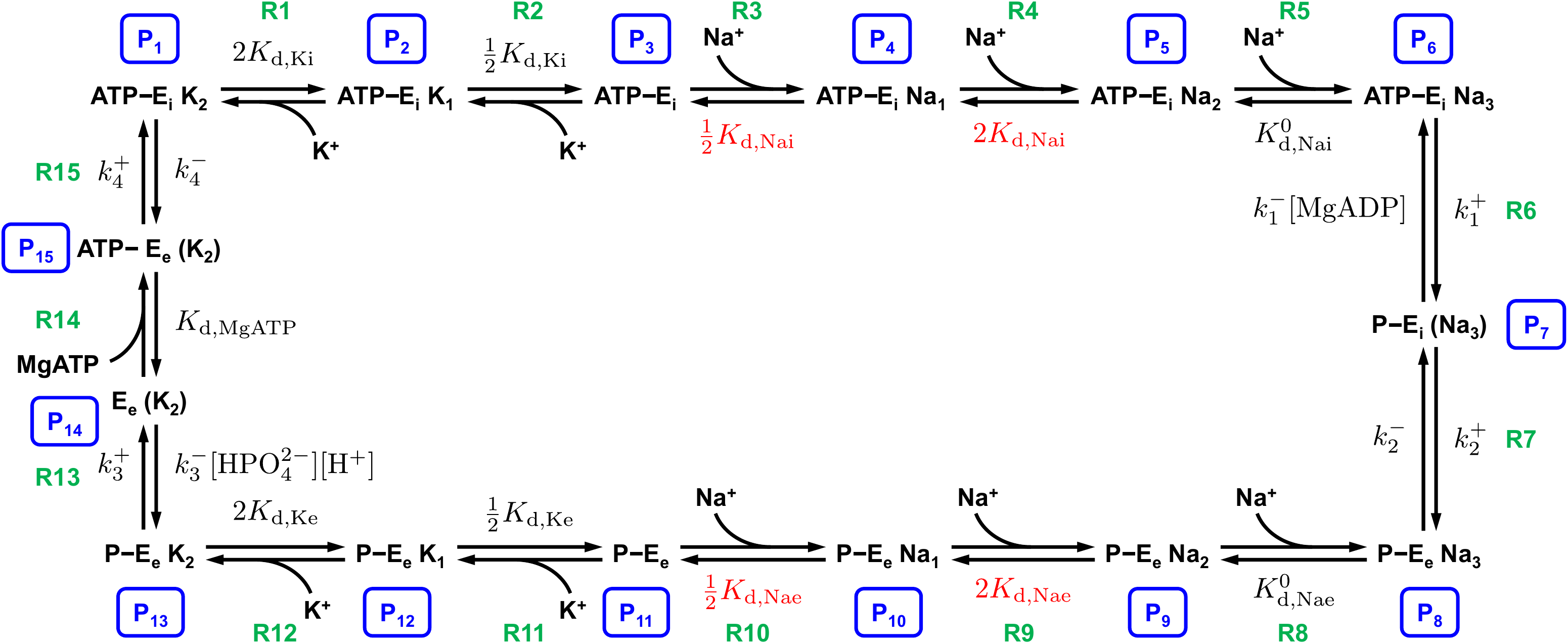}
\caption{\textbf{Reaction scheme of the cardiac Na$^+$/K$^+$ ATPase model.} Numbers for each pump state (\textit{blue boxes}) and reaction names (\textit{green}) are labelled, with corrected parameters shown in red.}
\label{fig:NaK_scheme}
\end{figure}

The Terkildsen \textit{et al.} model uses the Post-Albers cycle \citep{apell_electrogenic_1989}, a model in which sodium and potassium ions bind individually on one side of the membrane, and unbind on the other side (\autoref{fig:NaK_scheme}). The full Post-Albers cycle was simplified to reduce computational complexity by assuming that faster reactions are in rapid equilibrium, reducing the full 15-state model to a four-state model with eight modified rate constants \citep{smith_development_2004}. The entire cycle was then assumed to be in steady-state such that the model simplified to a single equation for cycling velocity, with metabolite dependence incorporated in a manner that accounted for thermodynamic constraints. We identified three issues while reimplementing the Terkildsen \textit{et al}. model, and made several modifications to remedy these issues:

\textbf{Issue 1:} Equilibrium constants were inconsistent with the number of binding sites. For identical binding sites the kinetic rate constants are typically assumed to be proportional to the number of sites available for binding/unbinding \citep{keener_mathematical_2009} and we modified the reaction scheme from \citet{terkildsen_balance_2007} to achieve this (\textit{red parameters within} \autoref{fig:NaK_scheme}).

\textbf{Issue 2:} The detailed balance constraint used during fitting procedure appears to have used an incorrect parameter value with important consequences on the thermodynamic consistency of the model. This constraint relates the kinetic constants defined in \autoref{fig:NaK_scheme}:
\begin{align}
	\frac{k_1^+ k_2^+ k_3^+ k_4^+ K_{d,\text{Na}_e}^0(K_{d,\text{Na}_e})^2 (K_{d,\text{K}_i})^2}
  {k_1^- k_2^- k_3^- k_4^- K_{d,\text{Na}_i}^0 (K_{d,\text{Na}_i})^2 (K_{d,\text{K}_e})^2 K_{d,\text{MgATP}}} = \exp\left( -\frac{\Delta G_\text{MgATP}^0}{RT} \right)
  \label{eq:detailed_balance}
\end{align}
where $R=8.314$ J/mol/K is the universal gas constant, $T$ is the absolute temperature, and $\Delta G_\mathrm{MgATP}^0$ is the standard free energy of MgATP hydrolysis at pH 0. It appears that \citet{terkildsen_balance_2007} started with a standard free energy of $-29.6\si{kJ/mol}$ at pH 7, but adjusted to a physiological pH rather than pH 0. As a result, substituting the model parameter values into equation \eqref{eq:detailed_balance} results in $\Delta G_\mathrm{MgATP}^0 = -30.2\si{kJ/mol}$, which is inconsistent with the typical standard free energy of $11.9\si{kJ/mol}$ at 311K \citep{tran_thermodynamic_2009,guynn_equilibrium_1973}. At a temperature of 310K this results in an overall equilibrium constant over $10^7$-fold greater than the correct value. Thus we use $\Delta G_\mathrm{MgATP}^0 = 11.9\si{kJ/mol}$ within the detailed balance constraint.

\textbf{Issue 3:} The lumping scheme used to reduce the 15-state model to a 4-state model with modified kinetic constants was similar to \citet{smith_development_2004}, however with the updated assignment of electrical dependence in \citet{terkildsen_balance_2007} some expressions were not applicable. As a result, expressions for several modified rate constants ($\alpha_1^+$, $\alpha_3^+$, $\alpha_2^-$ and $\alpha_4^-$) were incorrect, leading to inaccurate representations of pump kinetics. We have corrected the equations for these modified rate constants:
\begin{align}
	\alpha_1^+ &= \frac{k_1^+  \tilde{\text{Na}}_{i,1}\tilde{\text{Na}}_{i,2}^2}{\tilde{\text{Na}}_{i,1}\tilde{\text{Na}}_{i,2}^2 + (1 + \tilde{\text{Na}}_{i,2})^2 + (1 + \tilde{\text{K}}_i)^2 -1} \\
	\alpha_3^+ &= \frac{k_3^+ \tilde{\text{K}}_e^{2}}{\tilde{\text{Na}}_{e,1}\tilde{\text{Na}}_{e,2}^2 + (1 + \tilde{\text{Na}}_{e,2})^2 + (1 + \tilde{\text{K}}_e)^2 -1 } \\
	\alpha_2^-&= \frac{k_2^- \tilde{\text{Na}}_{e,1}\tilde{\text{Na}}_{e,2}^2}{\tilde{\text{Na}}_{e,1}\tilde{\text{Na}}_{e,2}^2 + (1 + \tilde{\text{Na}}_{e,2})^2 + (1 + \tilde{\text{K}}_e)^2 -1} \\
	\alpha_4^- &= \frac{k_4^- \tilde{\text{K}}_i^2}{\tilde{\text{Na}}_{i,1}\tilde{\text{Na}}_{i,2}^2 + (1 + \tilde{\text{Na}}_{i,2})^2 + (1 + \tilde{\text{K}}_i)^2 -1}
\end{align}
where
\begin{align}
	&\tilde{\text{Na}}_{i,1} = \frac{\mathrm{[Na^+]_i}}{K_{d,\text{Na}_i}^0 e^{\Delta FV/RT}} 
	&&\tilde{\text{Na}}_{i,2} = \frac{\mathrm{[Na^+]_i}}{K_{d,\text{Na}_i}} \\
	&\tilde{\text{Na}}_{e,1} = \frac{\mathrm{[Na^+]_e}}{K_{d,\text{Na}_e^0} e^{(1+\Delta) zFV/RT}} 
	&&\tilde{\text{Na}}_{e,2} = \frac{\mathrm{[Na^+]_e}}{K_{d,\text{Na}_e}} \\
	&\tilde{\text{K}}_i = \frac{\mathrm{[K^+]_i}}{K_{d,\text{K}_i}} 
	&&\tilde{\text{K}}_e = \frac{\mathrm{[K^+]_e}}{K_{d,\text{K}_e}} 
\end{align}
and $\Delta$ is the unit of charge translocated by the final sodium binding reaction R5. We derive an expression for $\alpha_1^+$, and expressions for the other modified rate constants follow similarly. Since the pump states 1 to 6 are lumped together, the constant $k_1^+$ is scaled according to the ratio between the amount of state 6 and the total amount of states 1--6. If we represent $x_i$ as the molar amount of state $i$, then:
\begin{align}
	\alpha_1^+ &= k_1^+ \frac{x_6}{x_6 + x_5 + x_4 + x_3 + x_2 + x_1} \notag \\
	&= k_1^+ \frac{1}{1 + x_5/x_6 + x_4/x_6 + x_3/x_6 + x_2/x_6 + x_1/x_6} \notag  \\
	&= \frac{k_1^+}{1 + 2\tilde{\text{Na}}_{i,1}^{-1} + 2\tilde{\text{Na}}_{i,1}^{-1}\tilde{\text{Na}}_{i,2}^{-1} + \tilde{\text{Na}}_{i,1}^{-1}\tilde{\text{Na}}_{i,2}^{-2} + 2\tilde{\text{Na}}_{i,1}^{-1}\tilde{\text{Na}}_{i,2}^{-2}\tilde{\text{K}}_i + \tilde{\text{Na}}_{i,1}^{-1}\tilde{\text{Na}}_{i,2}^{-2}\tilde{\text{K}}_i^2} \notag  \\
	&= \frac{k_1^+  \tilde{\text{Na}}_{i,1}\tilde{\text{Na}}_{i,2}^2}{\tilde{\text{Na}}_{i,1}\tilde{\text{Na}}_{i,2}^2 + (1 + \tilde{\text{Na}}_{i,2})^2 + (1 + \tilde{\text{K}}_i)^2 -1}
\end{align}

Because it was not possible to fix the above issues without significantly changing the kinetics of the model, we subsequently had to reparameterise the Terkildsen \textit{et al}. model such that it would be physically and thermodynamically consistent. In subsequent sections, we shall refer to the reparameterised model with updated equations as the ``updated model'' and the model with equations and parameters described in \citep{terkildsen_balance_2007} as the ``original model''.

\section{Reparameterisation of the model}
\label{sec:raparameterisation}
\begin{figure}
	\centering
	\begin{tabular}{c c}
		(a) & \multirow{4}{0.4\linewidth}[0cm]{
			\begin{minipage}{\linewidth}
				\vspace{2cm}
				\begin{align*}
				\mathrm{[Na^+]_i} &= 50\si{mM} \\
				\mathrm{[K^+]_i} &= 0\si{mM} \\
				\mathrm{[K^+]_e} &= 5.4\si{mM} \\
				\mathrm{pH} &= 7.4 \\
				\mathrm{[Pi]_{tot}} &= 0\si{mM} \\
				\mathrm{[MgATP]} &= 10\si{mM} \\
				\mathrm{[MgADP]} &= 0\si{mM} \\
				T &= 310\si{K}
				\end{align*}
		\end{minipage} } \\
		\includegraphics[width=0.4\linewidth]{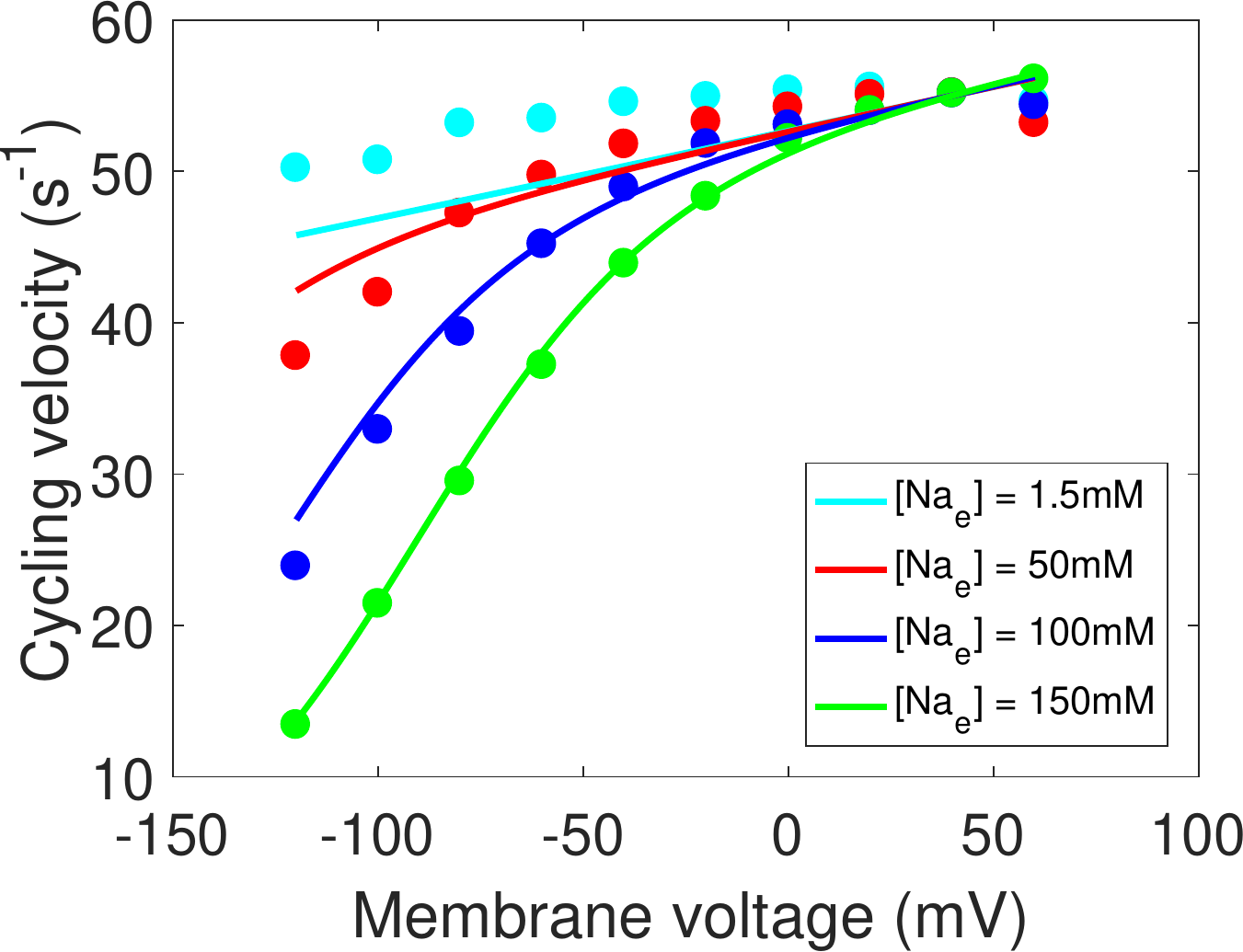} &
		\\
		(b) &   \\
		\includegraphics[width=0.4\linewidth]{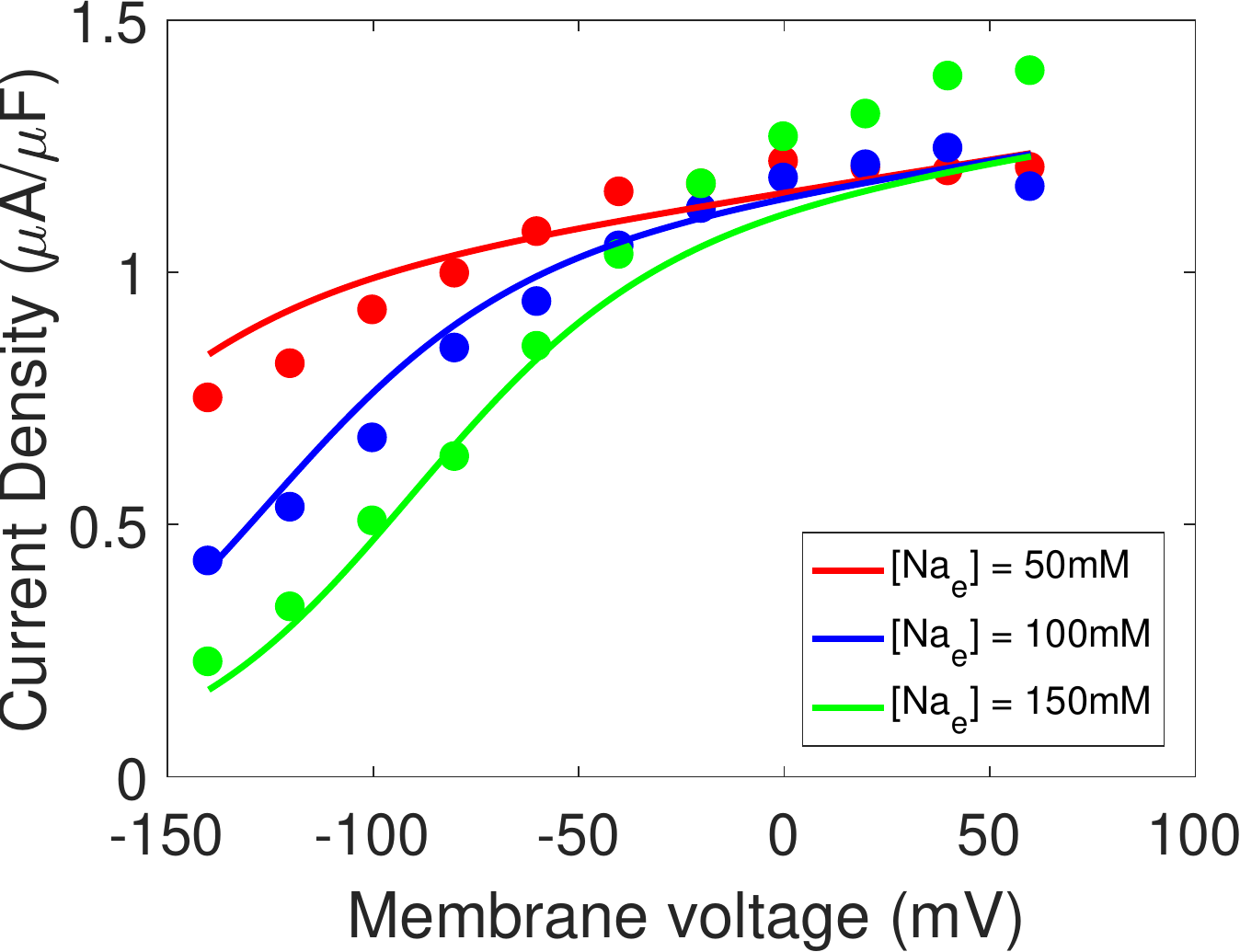} & 
	\end{tabular}
	\caption{\textbf{Model fit of the updated cardiac Na$^+$/K$^+$ ATPase model to current-voltage measurements.} \textbf{(a)} Comparison of the model to extracellular sodium and voltage data \citep[Fig. 3]{nakao_[na]_1989}, with cycling velocities scaled to a value of $55\si{s^{-1}}$ at $V = 40\si{mV}$. \textbf{(b)} Comparison of the model to whole-cell current measurements \citep[Fig. 2(a)]{nakao_[na]_1989}.} 
	\label{fig:fitting}
\end{figure}

\begin{figure}
	\centering
	\begin{tabular}{c c}
		(a) &  \multirow{2}{0.4\linewidth}[1cm]{
			\begin{minipage}{\linewidth}
				\small
				\begin{align*}
				V &= 0\si{mV}\\ 
				\mathrm{[Na^+]_e} &= 0\si{mM}\\ 
				\mathrm{[K^+]_i} &= 80\si{mM}\\ 
				\mathrm{[K^+]_e} &= 15\si{mM}\\ 
				\mathrm{pH} &= 7.2\\ 
				\mathrm{[Pi]_{tot}} &= 1\si{mM}\\ 
				\mathrm{[MgATP]} &= 2\si{mM}\\ 
				\mathrm{[MgADP]} &= 0\si{mM}\\ 
				T &= 308\si{K}
				\end{align*}
		\end{minipage}}\\
		\includegraphics[width=0.4\linewidth]{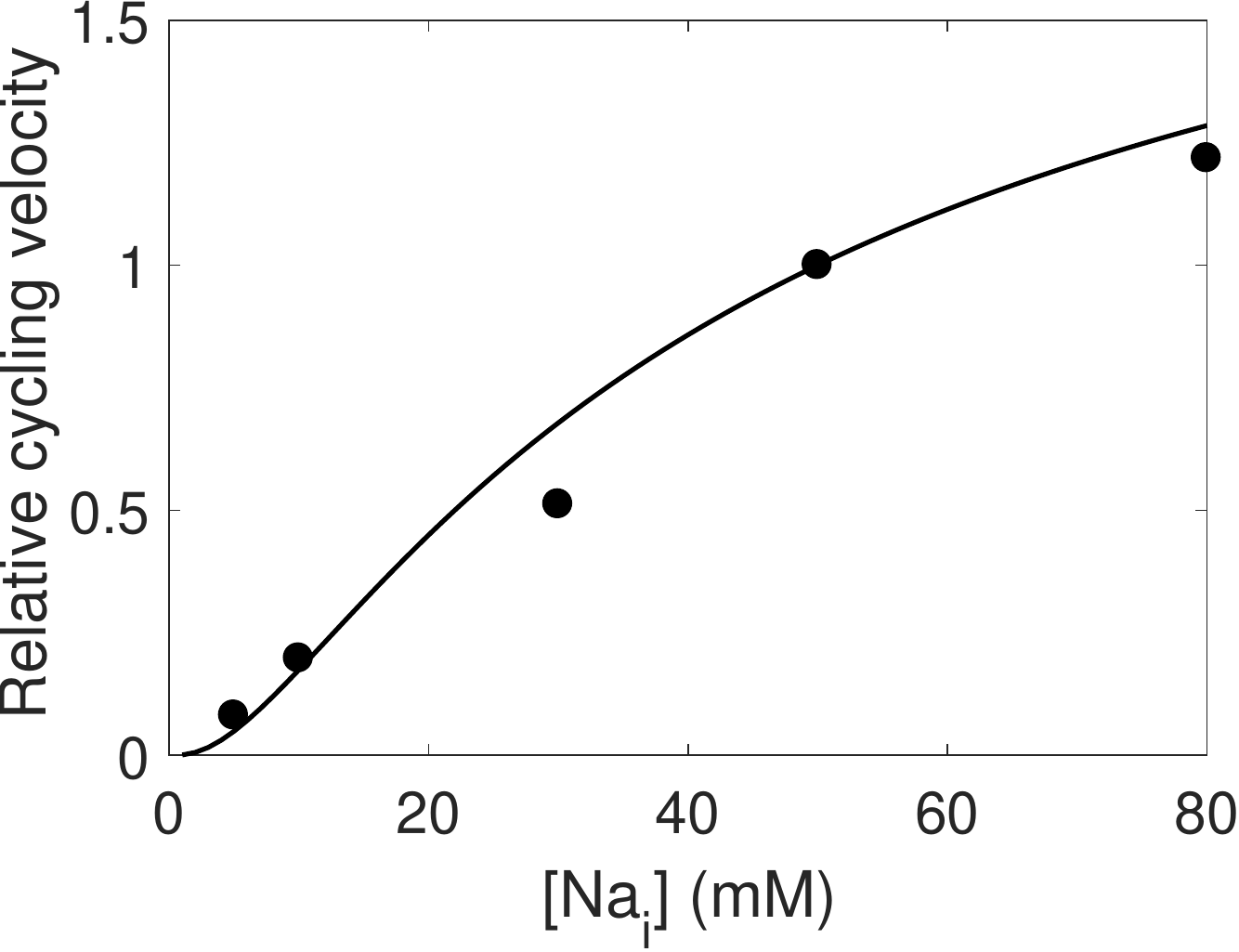} & \\[0.2cm]  
		(b) &  \multirow{2}{0.4\linewidth}[1cm]{
			\begin{minipage}{\linewidth}
				\small
				\begin{align*}
				V &= 0\si{mV}\\ 
				\mathrm{[Na^+]_i} &= 50\si{mM}\\ 
				\mathrm{[Na^+]_e} &= 150\si{mM}\\ 
				\mathrm{[K^+]_i} &= 140\si{mM}\\ 
				\mathrm{pH} &= 7.4\\ 
				\mathrm{[Pi]_{tot}} &= 0.5\si{mM}\\ 
				\mathrm{[MgATP]} &= 10\si{mM}\\ 
				\mathrm{[MgADP]} &= 0.02\si{mM}\\ 
				T &= 310\si{K}
				\end{align*}
		\end{minipage}}\\
		\includegraphics[width=0.4\linewidth]{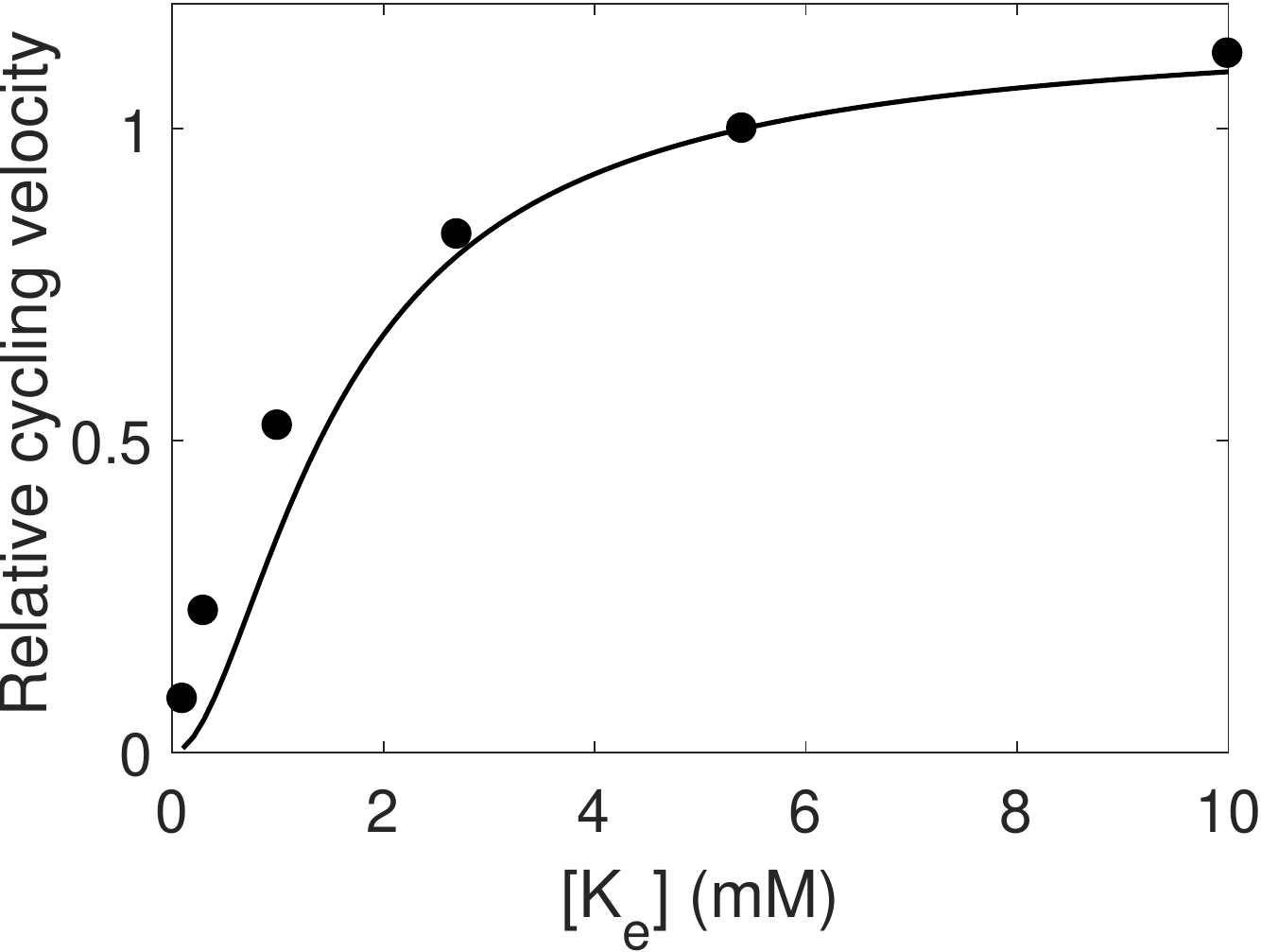} & \\[0.2cm]  
		(c) & \multirow{2}{0.4\linewidth}[1cm]{
			\begin{minipage}{\linewidth}
				\small
				\begin{align*}
				V &= 0\si{mV}\\ 
				\mathrm{[Na^+]_i} &= 40\si{mM}\\ 
				\mathrm{[Na^+]_e} &= 0\si{mM}\\ 
				\mathrm{[K^+]_i} &= 0\si{mM}\\ 
				\mathrm{[K^+]_e} &= 5\si{mM}\\ 
				\mathrm{pH} &= 7.4\\ 
				\mathrm{[Pi]_{tot}} &= 0\si{mM}\\ 
				\mathrm{[MgADP]} &= 0\si{mM}\\ 
				T &= 297\si{K}
				\end{align*}
		\end{minipage}}\\  
		\includegraphics[width=0.4\linewidth]{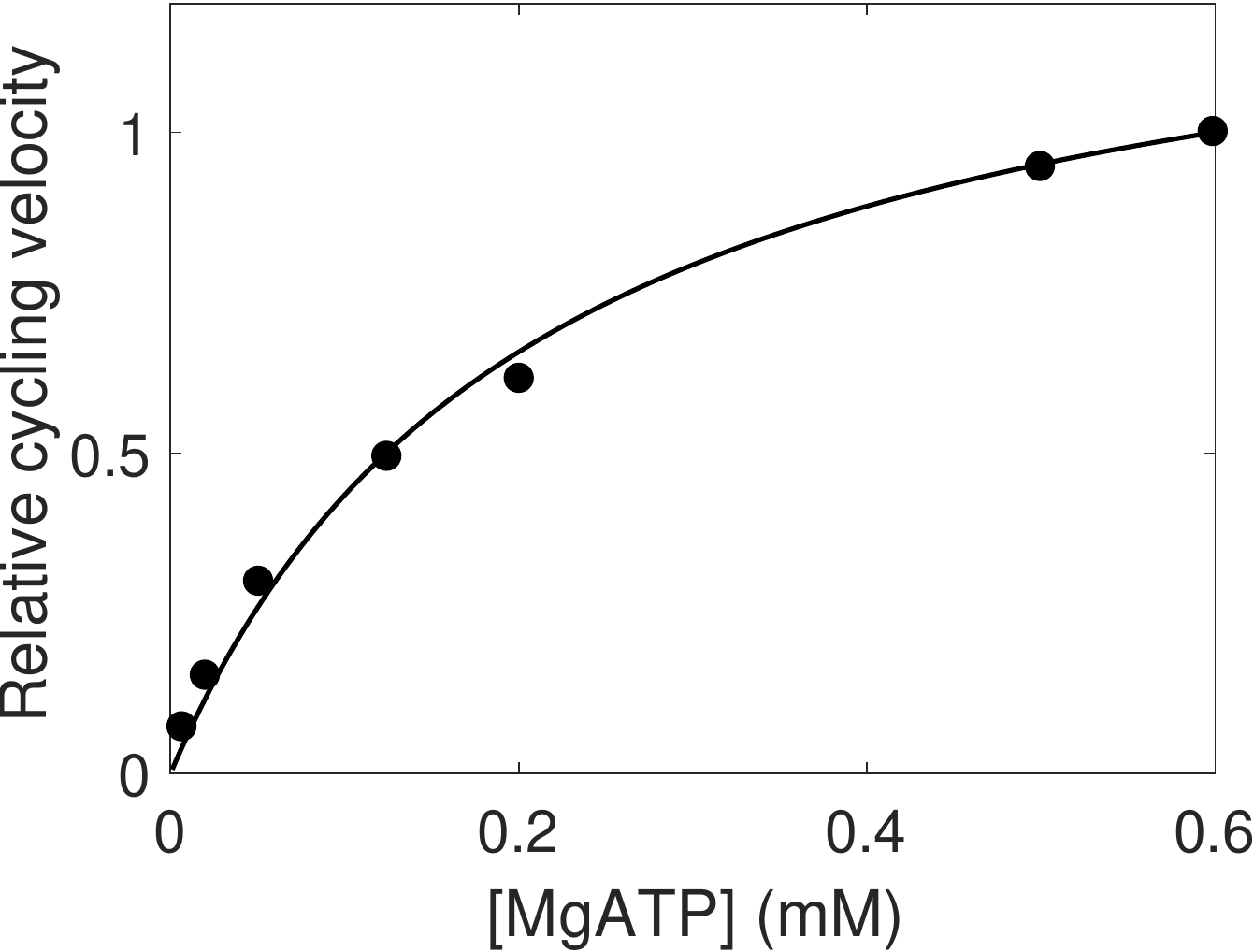} & 
	\end{tabular}
	\caption{\textbf{Model fit of the updated cardiac Na$^+$/K$^+$ ATPase model to metabolite dependence data.} Simulation conditions are displayed on the right of each figure. \textbf{(a)} Comparison of the model to data with varying intracellular sodium concentrations \citep[Fig. 7(a)]{hansen_dependence_2002}, normalised to the cycling velocity at $\mathrm{[Na]_i = 50\si{mM}}$. \textbf{(b)} Comparison of the model to data with varying extracellular potassium \citep[Fig. 11(a)]{nakao_[na]_1989}, normalised to the cycling velocity at $\mathrm{[K]_e = 5.4\si{mM}}$. \textbf{(c) } Comparison of the model to data with varying ATP \citep[Fig. 3(b)]{friedrich_na+k+-atpase_1996}, normalised to the cycling velocity at $\mathrm{[MgATP] = 0.6\si{mM}}$.}
	\label{fig:metabolite_dependence}
\end{figure}

Using the updated model's equations, we fitted parameters to data from Terkildsen \textit{et al.} \citep{terkildsen_balance_2007,terkildsen_modelling_2006}. After setting $\Delta G_\mathrm{MgATP}^0$ to $11.9\si{kJ/mol}$ in the thermodynamic constraint (Equation \eqref{eq:detailed_balance}), we parameterised the updated model using similar methods to the original model \citet{terkildsen_modelling_2006} which minimised an objective function describing divergence of the model from experimental data. Minor changes to the fitting procedure include:
\begin{enumerate}
	\item The weighting for extracellular potassium above 5.4 mM for the data of \citet{nakao_[na]_1989} was increased from $6\times$ to $15\times$ to obtain a reasonable fit at physiological concentrations.
	\item To ensure that cycling velocity magnitudes matched \citet{nakao_[na]_1989}, the curve for $[\mathrm{Na}]_e=150 \si{mM}$ was fitted without normalisation.
	\item Rather than using a local optimiser with literature sources for initial parameter estimates, we minimised the objective function by using particle swarm optimisation \citep{kennedy_particle_1995} followed by a local optimiser to find a global minimum.
\end{enumerate}
The updated model provides good fits to each data source (Figures \ref{fig:fitting} \& \ref{fig:metabolite_dependence}), and the quality of fits are comparable to \citet{terkildsen_modelling_2006}, although we achieved a slightly worse fit at lower extracellular sodium concentrations (\autoref{fig:fitting}(a)). It should be noted, however, that our model appears to be more consistent with experimental data that suggest saturated cycling velocity at positive membrane potentials is relatively insensitive to extracellular sodium (\autoref{fig:fitting}(b)) \citep{nakao_[na]_1989}. Updated model parameters are given in \autoref{tab:Terkildsen_parameters} of \autoref{sec:parameters}.

% * <ktra014@aucklanduni.ac.nz> 2017-09-22T02:43:11.081Z:
% 
% Figure 2a shows unnormalised cycling velocity (same as 2b) - should it be normalised?
% 
% ^ <panm@student.unimelb.edu.au> 2017-09-25T01:00:04.320Z:
% 
% The cycling velocity is "normalised", although I'm not sure if that's the right word as it's done in a different way from 2d-f. In Figure 2a, the curves were scaled so that they all passed through 55s^-1 at -40mV - which is how the model was compared to data in the original paper. I plot the cycling velocities without this scaling in Figure 2b, although it turns out there scaling didn't change the values by much. The scaling had a more noticeable impact in the original paper.
% 
% ^.

\begin{figure}
\centering
\begin{tabular}{c c c}
(a) & (b) & (c) \\
\includegraphics[width=0.3\linewidth]{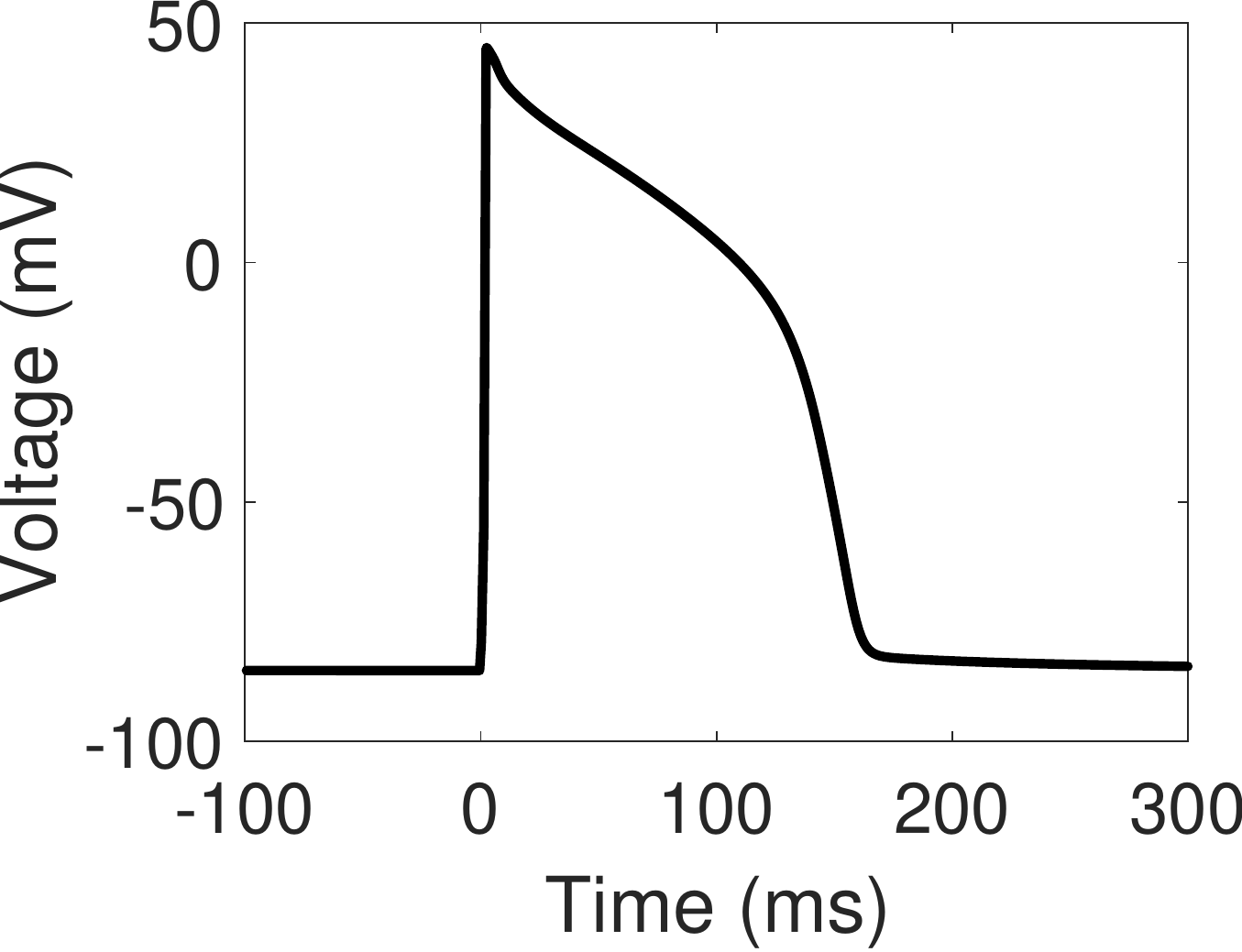} &
\includegraphics[width=0.3\linewidth]{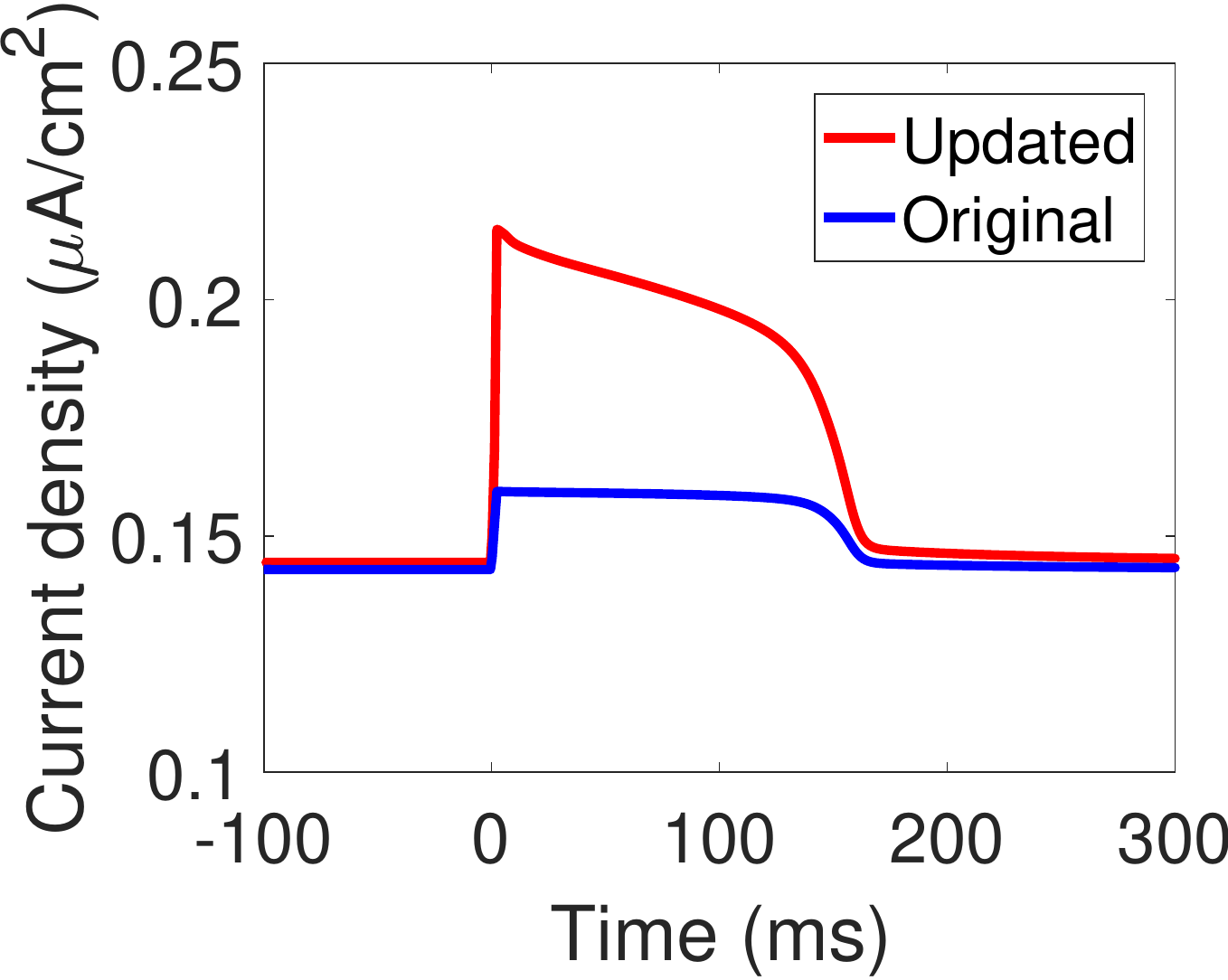} &
\includegraphics[width=0.3\linewidth]{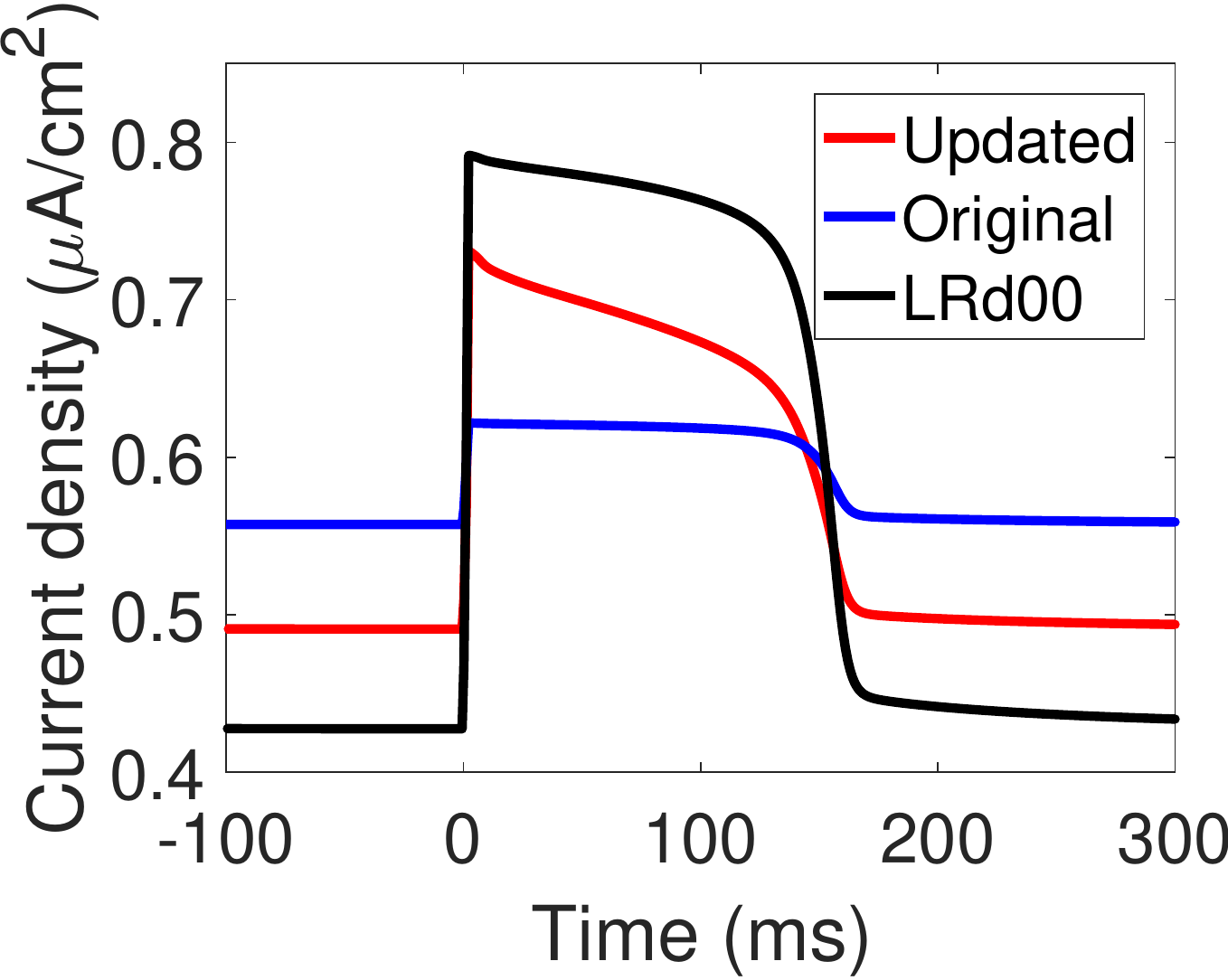} 
\end{tabular}
\caption{\textbf{A comparison of the updated cardiac Na$^+$/K$^+$ ATPase model to existing models.} (a) The action potential waveform used for pump simulation \citep{faber_action_2000}; (b) The Na$^+$/K$^+$ ATPase currents of the original and updated models; (c) A comparison of scaled versions of the updated and original models against the Na$^+$/K$^+$ ATPase model in \citet{faber_action_2000}. The pump density was increased by a factor of 3.4 in the updated model, and by a factor of 3.9 in the original model. $\mathrm{[Na^+]_i} = 10\si{mM}$, $\mathrm{[Na^+]_e} = 140\si{mM}$, $\mathrm{[K^+]_i} = 145\si{mM}$, $\mathrm{[K^+]_e} = 5.4\si{mM}$, $\mathrm{pH} = 7.095$, $\mathrm{[Pi]_{tot}} = 0.8\si{mM}$, $\mathrm{[MgATP]} = 6.95\si{mM}$, $\mathrm{[MgADP]} = 0.035\si{mM}$, $T = 310\si{K}$.}
\label{fig:updated_original_comp}
\end{figure}

The response of the updated model to an action potential input was simulated by using an action potential waveform generated from the Luo and Rudy (2000) model \citep{faber_action_2000,luo_dynamic_1994} (\autoref{fig:updated_original_comp}(a)). The original and updated models behave almost identically at resting membrane potentials, but the updated model has a much higher current during the action potential (\autoref{fig:updated_original_comp}(b)). As noted in \citet{terkildsen_modelling_2006}, the current of the pump is far lower at physiological intracellular sodium concentrations, thus the pump density needs to be appropriately scaled to be compatible with the Luo-Rudy model. Scaled versions of the Na$^+$/K$^+$ ATPase current within the updated and original models are qualitatively similar to that described using the Luo-Rudy equations \citep{faber_action_2000,luo_dynamic_1994}, however there are some differences in the resulting waveforms (\autoref{fig:updated_original_comp}(c)). In particular, the updated model behaves more similarly to the Luo-Rudy Na$^+$/K$^+$ ATPase formulation because it has a more variable current, and thus we hypothesise that under physiological concentrations, the updated model is more compatible with the whole-cell model by Luo and Rudy. A CellML version of the updated model is included with this manuscript.
% * <ktra014@aucklanduni.ac.nz> 2017-09-22T03:20:41.581Z:
% 
% I wonder if its worth using SED-ML to create settings for recreating all the figures.  But I assume your matlab code does all that already?
% 
% ^ <panm@student.unimelb.edu.au> 2017-09-25T01:30:52.320Z:
% 
% My MATLAB code can recreate the figures (although I haven't tested it on other computers). I haven't learned how to write SED-ML yet, but will look into it.
%
% ^.

\begin{figure}
	\centering
	\includegraphics[width=0.7\linewidth]{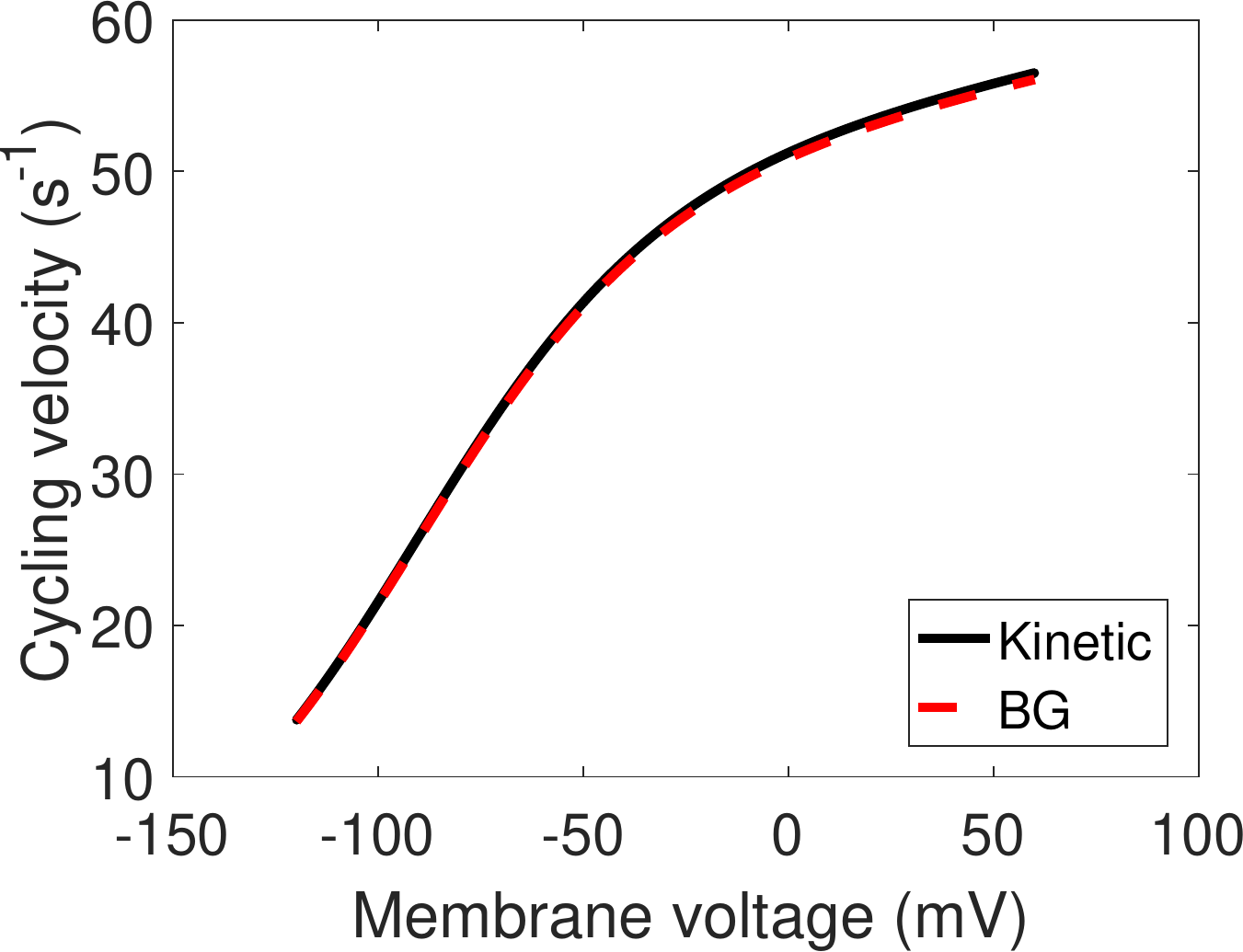}
	\caption{\textbf{A comparison of the kinetic and bond graph cardiac Na$^+$/K$^+$ models.} $\mathrm{[Na^+]_i} = 50\si{mM}$, $\mathrm{[Na^+]_e} = 150\si{mM}$, $\mathrm{[K^+]_i} = 0\si{mM}$, $\mathrm{[K^+]_e} = 5.4\si{mM}$, $\mathrm{pH} = 7.4$, $\mathrm{[Pi]_{tot}} = 0\si{mM}$, $\mathrm{[MgATP]} = 10\si{mM}$, $\mathrm{[MgADP]} = 0\si{mM}$, $T = 310\si{K}$.  The bond graph model is formulated using concentration ratios thus zero concentrations were approximated by a concentration of 0.001mM to avoid numerical errors.}
	\label{fig:kinetic_BG_comp}
\end{figure}

\section{Bond graph model}
To verify the physical plausibility of the updated model and to aid incorporation into larger models of cardiac energetics, we have also developed a bond graph version. Bond graphs are an energy-based approach to modelling physical systems, thus they ensure thermodynamic consistency \citep{gawthrop_energy-based_2014}. The structure of the bond graph model is given in \autoref{fig:Terkildsen_NaK} of \autoref{sec:BG_structure}. The process of converting the model into a bond graph required two notable changes to its representation. Firstly, the bond graph model represents the full unsimplified biochemical cycle, and reactions originally assumed to be in rapid equilibrium were replaced by reactions with fast kinetic parameters that conferred the same equilibrium constant. Thus the bond graph model contains 15 states, and is a close but not exact approximation of the kinetic model. Secondly, because kinetic parameters are often thermodynamically inconsistent \citep{liebermeister_modular_2010}, the bond graph approach requires chemical reaction networks to be specified using a different set of parameters: the reaction rate constant $\kappa$ and species thermodynamic constant $K$ \citep{gawthrop_energy-based_2014}. These parameters always describe thermodynamically consistent systems, regardless of their numerical value. As a result, we converted the kinetic parameters of our model into an equivalent set of bond graph parameters \citep{gawthrop_hierarchical_2015} by using the following matrix equation:
\begin{align}
	\textbf{Ln}(\mathbf{k}) = \mathbf{M} \textbf{Ln}(\mathbf{W} \boldsymbol{\lambda}) \label{eq:bg_general}
\end{align}
where $\textbf{Ln}$ is the element-wise logarithm operator. The vector $\mathbf{k}$ contains the kinetic parameters, $\boldsymbol{\lambda}$ contains the bond graph parameters, and $\mathbf{M}$ contains stoichiometric information. The partitions of these matrices are defined as:
\begin{align}
	\textbf{k} = \begin{bmatrix}
	k^+ \\ k^- \\ k^c
	\end{bmatrix}, \quad
	\textbf{M} = \left[ \begin{array}{c | c}
	I_{n_r \times n_r} & {N^f}^T \\ \hline
	I_{n_r \times n_r} & {N^r}^T \\ \hline
	0 & N^c
	\end{array} \right], \quad
	\boldsymbol{\lambda} = \begin{bmatrix}
	\kappa \\ K
	\end{bmatrix}
	\label{eq:standard}
\end{align}
where
\begin{align}
	k^+ &= \text{column vector of forward rate constants} \\
	k^- &= \text{column vector of reverse rate constants} \\
	\kappa &= \text{column vector of reaction rate constants} \\
	K &= \text{column vector of species thermodynamic constants} \\
	N^f &= \text{forward stoichiometric matrix} \\
	N^r &= \text{reverse stoichiometric matrix}
\end{align}
The matrices $k^c$ and $N^c$ were used to enforce further constraints between the thermodynamic constants of different species, in particular the equilibria of individual ions present within different compartments, and the equilibrium constant of ATP hydrolysis. Assuming that equation \eqref{eq:bg_general} can be solved, one possible solution is given by
\begin{align}
	\boldsymbol{\lambda_0 } = \mathbf{W}^{-1} \textbf{Exp} (\mathbf{M}^\dagger \textbf{Ln} (\mathbf{k}))
\end{align}
where $\textbf{Exp}$ is the element-wise exponential operator and $\mathbf{M}^\dagger$ is the Moore-Penrose pseudo-inverse of $\mathbf{M}$. Since the bond graph framework is energy-based, species must be expressed as molar amounts rather than concentrations to adequately compare energies in different compartments. Therefore we use the diagonal matrix $\mathbf{W}$ to scale the species thermodynamic constants according to the volume of the compartments they reside in. For consistency with \citet{terkildsen_balance_2007}, an intracellular volume of $W_i = 38\si{pL}$ was used for the species $\mathrm{Na_i^+}$, $\mathrm{K_i^+}$, $\mathrm{MgATP}$, $\mathrm{MgADP}$, $\mathrm{Pi}$ and $\mathrm{H^+}$, an extracellular volume of $W_e = 5.182\si{pL}$ was used for $\mathrm{Na_e^+}$, $\mathrm{K_e^+}$, and a constant of 1 was used for each of the pump states.

The bond graph model was simulated under the conditions described in \autoref{fig:fitting}(a) to reproduce the curve for $\mathrm{[Na^+]_e}=150\si{mM}$, using a slowly increasing membrane voltage to induce quasi-steady-state behaviour. There is a high degree of correspondence between the kinetic and bond graph models (\autoref{fig:kinetic_BG_comp}). The closeness of the two models suggests that the fast kinetic constants are good approximations of reactions in rapid equilibrium, although we note that the deviation between the bond graph and kinetic models increases slightly at higher cycling velocities when the faster reactions begin to limit flux through the cycle. CellML code describing the bond graph model and reproducing the curve in \autoref{fig:kinetic_BG_comp} is provided with this manuscript.

\section{Conclusion}
In this manuscript we describe an updated model for the cardiac Na$^+$/K$^+$ ATPase model originally developed by \citet{terkildsen_balance_2007}. We have corrected errors with the original model formulation and refitted necessary parameters to ensure that the resulting model is thermodynamically consistent while still recapitulating a wide range of experimental data. We note that the updated model has a natural bond graph representation, and include CellML and MATLAB code for both the kinetic and bond graph models to aid reproducibility. We believe that the thermodynamic consistency and improved reusability of our updated model make it ideal for incorporation into future whole-cell models to study cardiac cell energetics.

\section{Acknowledgements}
This research was supported in part by the Australian Government through the Australian Research Council's Discovery Projects funding scheme (project DP170101358), an Australian Government Research Training Program Scholarship to M.P., and a Research Fellowship (1692) from the Heart Foundation of New Zealand to K.T.

\bibliographystyle{model2-names_edit}
\small
\bibliography{bibliography}{}

\begin{thebibliography}{21}
\expandafter\ifx\csname natexlab\endcsname\relax\def\natexlab#1{#1}\fi
\providecommand{\url}[1]{\texttt{#1}}
\providecommand{\href}[2]{#2}
\providecommand{\path}[1]{#1}
\providecommand{\DOIprefix}{doi:}
\providecommand{\ArXivprefix}{arXiv:}
\providecommand{\URLprefix}{URL: }
\providecommand{\Pubmedprefix}{pmid:}
\providecommand{\doi}[1]{\href{http://dx.doi.org/#1}{\path{#1}}}
\providecommand{\Pubmed}[1]{\href{pmid:#1}{\path{#1}}}
\providecommand{\bibinfo}[2]{#2}
\ifx\xfnm\relax \def\xfnm[#1]{\unskip,\space#1}\fi
%Type = Article
\bibitem[{Apell(1989)}]{apell_electrogenic_1989}
\bibinfo{author}{Apell, H.J.}, \bibinfo{year}{1989}.
\newblock \bibinfo{title}{Electrogenic properties of the {Na},{K} pump}.
\newblock \bibinfo{journal}{The Journal of Membrane Biology}
  \bibinfo{volume}{110}, \bibinfo{pages}{103--114}.
%Type = Book
\bibitem[{Borutzky(2010)}]{borutzky_bond_2010}
\bibinfo{author}{Borutzky, W.}, \bibinfo{year}{2010}.
\newblock \bibinfo{title}{Bond {Graph} {Methodology}}.
\newblock \bibinfo{publisher}{Springer}.
%Type = Article
\bibitem[{Faber and Rudy(2000)}]{faber_action_2000}
\bibinfo{author}{Faber, G.M.}, \bibinfo{author}{Rudy, Y.},
  \bibinfo{year}{2000}.
\newblock \bibinfo{title}{Action {Potential} and {Contractility} {Changes} in
  [{Na}$^+$]\textsubscript{i} {Overloaded} {Cardiac} {Myocytes}: {A}
  {Simulation} {Study}}.
\newblock \bibinfo{journal}{Biophysical Journal} \bibinfo{volume}{78},
  \bibinfo{pages}{2392--2404}.
%Type = Article
\bibitem[{Friedrich et~al.(1996)Friedrich, Bamberg and
  Nagel}]{friedrich_na+k+-atpase_1996}
\bibinfo{author}{Friedrich, T.}, \bibinfo{author}{Bamberg, E.},
  \bibinfo{author}{Nagel, G.}, \bibinfo{year}{1996}.
\newblock \bibinfo{title}{Na$^+$,{K}$^+$-{ATPase} pump currents in giant
  excised patches activated by an {ATP} concentration jump}.
\newblock \bibinfo{journal}{Biophysical Journal} \bibinfo{volume}{71},
  \bibinfo{pages}{2486--2500}.
%Type = Article
\bibitem[{Gawthrop and Bevan(2007)}]{gawthrop_bond-graph_2007}
\bibinfo{author}{Gawthrop, P.}, \bibinfo{author}{Bevan, G.},
  \bibinfo{year}{2007}.
\newblock \bibinfo{title}{Bond-graph modeling}.
\newblock \bibinfo{journal}{IEEE Control Systems} \bibinfo{volume}{27},
  \bibinfo{pages}{24--45}.
%Type = Book
\bibitem[{Gawthrop and Smith(1996)}]{gawthrop_metamodelling:_1996}
\bibinfo{author}{Gawthrop, P.}, \bibinfo{author}{Smith, L.},
  \bibinfo{year}{1996}.
\newblock \bibinfo{title}{Metamodelling: for bond graphs and dynamic systems}.
\newblock Prentice {Hall} international series in systems and control
  engineering, \bibinfo{publisher}{Prentice Hall}, \bibinfo{address}{London,
  New York}.
%Type = Article
\bibitem[{Gawthrop and Crampin(2014)}]{gawthrop_energy-based_2014}
\bibinfo{author}{Gawthrop, P.J.}, \bibinfo{author}{Crampin, E.J.},
  \bibinfo{year}{2014}.
\newblock \bibinfo{title}{Energy-based analysis of biochemical cycles using
  bond graphs}.
\newblock \bibinfo{journal}{Proceedings of the Royal Society of London A:
  Mathematical, Physical and Engineering Sciences} \bibinfo{volume}{470},
  \bibinfo{pages}{20140459}.
%Type = Article
\bibitem[{Gawthrop et~al.(2015)Gawthrop, Cursons and
  Crampin}]{gawthrop_hierarchical_2015}
\bibinfo{author}{Gawthrop, P.J.}, \bibinfo{author}{Cursons, J.},
  \bibinfo{author}{Crampin, E.J.}, \bibinfo{year}{2015}.
\newblock \bibinfo{title}{Hierarchical bond graph modelling of biochemical
  networks}.
\newblock \bibinfo{journal}{Proc. R. Soc. A} \bibinfo{volume}{471},
  \bibinfo{pages}{20150642}.
%Type = Article
\bibitem[{Guynn and Veech(1973)}]{guynn_equilibrium_1973}
\bibinfo{author}{Guynn, R.W.}, \bibinfo{author}{Veech, R.L.},
  \bibinfo{year}{1973}.
\newblock \bibinfo{title}{The {Equilibrium} {Constants} of the {Adenosine}
  {Triphosphate} {Hydrolysis} and the {Adenosine} {Triphosphate}-{Citrate}
  {Lyase} {Reactions}}.
\newblock \bibinfo{journal}{Journal of Biological Chemistry}
  \bibinfo{volume}{248}, \bibinfo{pages}{6966--6972}.
%Type = Article
\bibitem[{Hansen et~al.(2002)Hansen, Buhagiar, Kong, Clarke, Gray and
  Rasmussen}]{hansen_dependence_2002}
\bibinfo{author}{Hansen, P.S.}, \bibinfo{author}{Buhagiar, K.A.},
  \bibinfo{author}{Kong, B.Y.}, \bibinfo{author}{Clarke, R.J.},
  \bibinfo{author}{Gray, D.F.}, \bibinfo{author}{Rasmussen, H.H.},
  \bibinfo{year}{2002}.
\newblock \bibinfo{title}{Dependence of {Na}$^+$-{K}$^+$ pump current-voltage
  relationship on intracellular {Na}$^+$, {K}$^+$, and {Cs}$^+$ in rabbit
  cardiac myocytes}.
\newblock \bibinfo{journal}{American Journal of Physiology - Cell Physiology}
  \bibinfo{volume}{283}, \bibinfo{pages}{C1511--C1521}.
%Type = Book
\bibitem[{Keener and Sneyd(2009)}]{keener_mathematical_2009}
\bibinfo{author}{Keener, J.}, \bibinfo{author}{Sneyd, J.},
  \bibinfo{year}{2009}.
\newblock \bibinfo{title}{Mathematical {Physiology}}. volume
  \bibinfo{volume}{8/1} of \textit{\bibinfo{series}{Interdisciplinary {Applied}
  {Mathematics}}}.
\newblock \bibinfo{publisher}{Springer New York}, \bibinfo{address}{New York,
  NY}.
%Type = Inproceedings
\bibitem[{Kennedy and Eberhart(1995)}]{kennedy_particle_1995}
\bibinfo{author}{Kennedy, J.}, \bibinfo{author}{Eberhart, R.},
  \bibinfo{year}{1995}.
\newblock \bibinfo{title}{Particle swarm optimization}, in:
  \bibinfo{booktitle}{{IEEE} {International} {Conference} on {Neural}
  {Networks}, 1995. {Proceedings}}, pp. \bibinfo{pages}{1942--1948 vol.4}.
%Type = Article
\bibitem[{Liebermeister et~al.(2010)Liebermeister, Uhlendorf and
  Klipp}]{liebermeister_modular_2010}
\bibinfo{author}{Liebermeister, W.}, \bibinfo{author}{Uhlendorf, J.},
  \bibinfo{author}{Klipp, E.}, \bibinfo{year}{2010}.
\newblock \bibinfo{title}{Modular rate laws for enzymatic reactions:
  thermodynamics, elasticities and implementation}.
\newblock \bibinfo{journal}{Bioinformatics} \bibinfo{volume}{26},
  \bibinfo{pages}{1528--1534}.
%Type = Article
\bibitem[{Lloyd et~al.(2004)Lloyd, Halstead and Nielsen}]{lloyd_cellml:_2004}
\bibinfo{author}{Lloyd, C.M.}, \bibinfo{author}{Halstead, M.D.B.},
  \bibinfo{author}{Nielsen, P.F.}, \bibinfo{year}{2004}.
\newblock \bibinfo{title}{{CellML}: its future, present and past}.
\newblock \bibinfo{journal}{Progress in Biophysics and Molecular Biology}
  \bibinfo{volume}{85}, \bibinfo{pages}{433--450}.
%Type = Article
\bibitem[{Luo and Rudy(1994)}]{luo_dynamic_1994}
\bibinfo{author}{Luo, C.H.}, \bibinfo{author}{Rudy, Y.}, \bibinfo{year}{1994}.
\newblock \bibinfo{title}{A dynamic model of the cardiac ventricular action
  potential. {I}. {Simulations} of ionic currents and concentration changes.}
\newblock \bibinfo{journal}{Circulation Research} \bibinfo{volume}{74},
  \bibinfo{pages}{1071--1096}.
%Type = Article
\bibitem[{Nakao and Gadsby(1989)}]{nakao_[na]_1989}
\bibinfo{author}{Nakao, M.}, \bibinfo{author}{Gadsby, D.C.},
  \bibinfo{year}{1989}.
\newblock \bibinfo{title}{[{Na}] and [{K}] dependence of the {Na}/{K} pump
  current-voltage relationship in guinea pig ventricular myocytes.}
\newblock \bibinfo{journal}{The Journal of General Physiology}
  \bibinfo{volume}{94}, \bibinfo{pages}{539--565}.
%Type = Article
\bibitem[{Oster et~al.(1971)Oster, Perelson and
  Katchalsky}]{oster_network_1971}
\bibinfo{author}{Oster, G.}, \bibinfo{author}{Perelson, A.},
  \bibinfo{author}{Katchalsky, A.}, \bibinfo{year}{1971}.
\newblock \bibinfo{title}{Network thermodynamics}.
\newblock \bibinfo{journal}{Nature} \bibinfo{volume}{234},
  \bibinfo{pages}{393--399}.
%Type = Article
\bibitem[{Smith and Crampin(2004)}]{smith_development_2004}
\bibinfo{author}{Smith, N.P.}, \bibinfo{author}{Crampin, E.J.},
  \bibinfo{year}{2004}.
\newblock \bibinfo{title}{Development of models of active ion transport for
  whole-cell modelling: cardiac sodium–potassium pump as a case study}.
\newblock \bibinfo{journal}{Progress in Biophysics and Molecular Biology}
  \bibinfo{volume}{85}, \bibinfo{pages}{387--405}.
%Type = Phdthesis
\bibitem[{Terkildsen(2006)}]{terkildsen_modelling_2006}
\bibinfo{author}{Terkildsen, J.}, \bibinfo{year}{2006}.
\newblock \bibinfo{title}{Modelling {Extracellular} {Potassium} {Accumulation}
  in {Cardiac} {Ischaemia}}.
\newblock \bibinfo{type}{Masters {Thesis}}. The University of Auckland.
%Type = Article
\bibitem[{Terkildsen et~al.(2007)Terkildsen, Crampin and
  Smith}]{terkildsen_balance_2007}
\bibinfo{author}{Terkildsen, J.R.}, \bibinfo{author}{Crampin, E.J.},
  \bibinfo{author}{Smith, N.P.}, \bibinfo{year}{2007}.
\newblock \bibinfo{title}{The balance between inactivation and activation of
  the {Na}$^+$-{K}$^+$ pump underlies the triphasic accumulation of
  extracellular {K}$^+$ during myocardial ischemia}.
\newblock \bibinfo{journal}{American Journal of Physiology - Heart and
  Circulatory Physiology} \bibinfo{volume}{293}, \bibinfo{pages}{H3036--H3045}.
%Type = Article
\bibitem[{Tran et~al.(2009)Tran, Smith, Loiselle and
  Crampin}]{tran_thermodynamic_2009}
\bibinfo{author}{Tran, K.}, \bibinfo{author}{Smith, N.P.},
  \bibinfo{author}{Loiselle, D.S.}, \bibinfo{author}{Crampin, E.J.},
  \bibinfo{year}{2009}.
\newblock \bibinfo{title}{A {Thermodynamic} {Model} of the {Cardiac}
  {Sarcoplasmic}/{Endoplasmic} {Ca}$^{2+}$ ({SERCA}) {Pump}}.
\newblock \bibinfo{journal}{Biophysical Journal} \bibinfo{volume}{96},
  \bibinfo{pages}{2029--2042}.

\end{thebibliography}
\normalsize

\appendix
\section{Parameters}
\label{sec:parameters}
\begin{table}[H]
	\caption{\textbf{Kinetic parameters for the updated cardiac Na$^+$/K$^+$ ATPase model.} Refer to \autoref{fig:NaK_scheme} for a schematic.}
	\centering
	\bgroup
	\def\arraystretch{1.3}
	\begin{tabular}{c p{0.48\linewidth} l}
		\toprule
		\textbf{Parameter} & \textbf{Description}& \textbf{Value} \\ \midrule
		$k_1^+$ & Forward rate constant of reaction R6 & $1423.2\ \si{s^{-1}}$\\ 
		$k_1^-$ & Reverse rate constant of reaction R6 & $225.9048\ \si{s^{-1}}$\\ 
		$k_2^+$ & Forward rate constant of reaction R7 & $11564.8064\ \si{s^{-1}}$\\ 
		$k_2^-$ & Reverse rate constant of reaction R7 & $36355.3201\ \si{s^{-1}}$\\ 
		$k_3^+$ & Forward rate constant of reaction R13 & $194.4506\ \si{s^{-1}}$\\ 
		$k_3^-$ & Reverse rate constant of reaction R13 & $281037.2758\ \si{mM^{-2}s^{-1}}$\\ 
		$k_4^+$ & Forward rate constant of reaction R15 & $30629.8836\ \si{s^{-1}}$\\ 
		$k_4^-$ & Reverse rate constant of reaction R15 & $1.574\times 10^{6}\ \si{s^{-1}}$\\ 
		$K_{\text{d,Nai}}^0$ & Voltage-dependent dissociation constant of intracellular $\mathrm{Na^+}$  & $579.7295\ \si{mM}$\\ 
		$K_{\text{d,Nae}}^0$ & Voltage-dependent dissociation constant of extracellular $\mathrm{Na^+}$ & $0.034879\ \si{mM}$\\ 
		$K_{\text{d,Nai}}$ & Voltage-independent dissociation constant of intracellular $\mathrm{Na^+}$ & $5.6399\ \si{mM}$\\ 
		$K_{\text{d,Nae}}$ & Voltage-independent dissociation constant of extracellular $\mathrm{Na^+}$  & $10616.9377\ \si{mM}$\\ 
		$K_{\text{d,Ki}}$ & Dissociation constant of intracellular $\mathrm{K^+}$ & $16794.976\ \si{mM}$\\ 
		$K_{\text{d,Ke}}$ & Dissociation constant of extracellular $\mathrm{K^+}$ & $1.0817\ \si{mM}$\\ 
		$K_{\text{d,MgATP}}$ & Dissociation constant of MgATP & $140.3709\ \si{mM}$\\ 
		$\Delta$ & Charge translocated by reaction R5 & $-0.0550$ \\
		Pump density & Number of pumps per $\si{\mu m^2}$  & $1360.2624\ \si{\mu m^{-2}}$  \\ \bottomrule& 
	\end{tabular}
	\egroup
	\label{tab:Terkildsen_parameters}
\end{table}

\begin{table}[H]
	\caption{\textbf{Parameters for the bond graph version of the updated cardiac Na$^+$/K$^+$ ATPase model.} Parameters were derived by using an intracellular volume of 38pL and an extracellular volume of 5.182pL. Refer to \autoref{fig:Terkildsen_NaK} for the bond graph schematic.}
	\centering
	\begin{tabular}{cl c l}
		\toprule
		\textbf{Component} & \textbf{Description}& \textbf{Parameter} & \textbf{Value} \\ \midrule
		R1 & Reaction R1 & $\kappa_1$ & $ 330.5462\ \si{fmol/s} $\\ 
		R2 & Reaction R2 & $\kappa_2$ & $ 132850.9145\ \si{fmol/s} $\\ 
		R3 & Reaction R3 & $\kappa_3$ & $ 200356.0223\ \si{fmol/s} $\\ 
		R4 & Reaction R4 & $\kappa_4$ & $ 2238785.3951\ \si{fmol/s} $\\ 
		R5 & Reaction R5 & $\kappa_5$ & $ 10787.9052\ \si{fmol/s} $\\ 
		R6 & Reaction R6 & $\kappa_6$ & $ 15.3533\ \si{fmol/s} $\\ 
		R7 & Reaction R7 & $\kappa_7$ & $ 2.3822\ \si{fmol/s} $\\ 
		R8 & Reaction R8 & $\kappa_8$ & $ 2.2855\ \si{fmol/s} $\\ 
		R9 & Reaction R9 & $\kappa_9$ & $ 1540.1349\ \si{fmol/s} $\\ 
		R10 & Reaction R10 & $\kappa_{10}$ & $ 259461.6507\ \si{fmol/s} $\\ 
		R11 & Reaction R11 & $\kappa_{11}$ & $ 172042.3334\ \si{fmol/s} $\\ 
		R12 & Reaction R12 & $\kappa_{12}$ & $ 6646440.3909\ \si{fmol/s} $\\ 
		R13 & Reaction R13 & $\kappa_{13}$ & $ 597.4136\ \si{fmol/s} $\\ 
		R14 & Reaction R14 & $\kappa_{14}$ & $ 70.9823\ \si{fmol/s} $\\ 
		R15 & Reaction R15 & $\kappa_{15}$ & $ 0.015489\ \si{fmol/s} $\\ 
		$\text{P}_1$ & Pump state ATP--E\textsubscript{i} K\textsubscript{2}
		 & $K_1$ & $101619537.2009\ \si{fmol^{-1}}$ \\ 
		$\text{P}_2$ & Pump state ATP--E\textsubscript{i} K\textsubscript{1}
		 & $K_2$ & $63209.8623\ \si{fmol^{-1}}$ \\ 
		$\text{P}_3$ & Pump state ATP--E\textsubscript{i}
		 & $K_3$ & $157.2724\ \si{fmol^{-1}}$ \\ 
		$\text{P}_4$ & Pump state ATP--E\textsubscript{i} Na\textsubscript{1}
		 & $K_4$ & $14.0748\ \si{fmol^{-1}}$ \\ 
		$\text{P}_5$ & Pump state ATP--E\textsubscript{i} Na\textsubscript{2}
		 & $K_5$ & $5.0384\ \si{fmol^{-1}}$ \\ 
		$\text{P}_6$ & Pump state ATP--E\textsubscript{i} Na\textsubscript{3}
		 & $K_6$ & $92.6964\ \si{fmol^{-1}}$ \\ 
		$\text{P}_7$ & Pump state P--E\textsubscript{i} (Na\textsubscript{3})
		 & $K_7$ & $4854.5924\ \si{fmol^{-1}}$ \\ 
		$\text{P}_8$ & Pump state P--E\textsubscript{e} Na\textsubscript{3}
		 & $K_8$ & $15260.9786\ \si{fmol^{-1}}$ \\ 
		$\text{P}_9$ & Pump state P--E\textsubscript{e} Na\textsubscript{2}
		 & $K_9$ & $13787022.8009\ \si{fmol^{-1}}$ \\ 
		$\text{P}_{10}$ & Pump state P--E\textsubscript{e} Na\textsubscript{1}
		 & $K_{10}$ & $20459.5509\ \si{fmol^{-1}}$ \\ 
		$\text{P}_{11}$ & Pump state P--E\textsubscript{e}
		 & $K_{11}$ & $121.4456\ \si{fmol^{-1}}$ \\ 
		$\text{P}_{12}$ & Pump state P--E\textsubscript{e} K\textsubscript{1}
		 & $K_{12}$ & $3.1436\ \si{fmol^{-1}}$ \\ 
		$\text{P}_{13}$ & Pump state P--E\textsubscript{e} K\textsubscript{2}
		 & $K_{13}$ & $0.32549\ \si{fmol^{-1}}$ \\ 
		$\text{P}_{14}$ & Pump state E\textsubscript{e} (K\textsubscript{2})
		& $K_{14}$ & $156.3283\ \si{fmol^{-1}}$ \\ 
		$\text{P}_{15}$ & Pump state ATP--Ee (K\textsubscript{2})
		& $K_{15}$ & $1977546.8577\ \si{fmol^{-1}}$ \\ 
		Ki & Intracellular $\text{K}_\text{i}^+$ & $K_\text{Ki}$ & $0.0012595\ \si{fmol^{-1}}$ \\ 
		Ke & Extracellular $\text{K}_\text{e}^+$ & $K_\text{Ke}$ & $0.009236\ \si{fmol^{-1}}$ \\ 
		Nai & Intracellular $\text{Na}_\text{i}^+$ & $K_\text{Nai}$ & $0.00083514\ \si{fmol^{-1}}$ \\ 
		Nae & Extracellular $\text{Na}_\text{e}^+$ & $K_\text{Nae}$ & $0.0061242\ \si{fmol^{-1}}$ \\ 
		$\text{MgATP}$ & Intracellular MgATP & $K_\text{MgATP}$ & $2.3715\ \si{fmol^{-1}}$ \\ 
		$\text{MgADP}$ & Intracellular MgADP & $K_\text{MgADP}$ & $7.976 \times 10^{-5} \ \si{fmol^{-1}}$ \\ 
		$\text{P}_\text{i}$ & Free inorganic phosphate & $K_\text{Pi}$ & $0.04565\ \si{fmol^{-1}}$ \\ 
		H & Intracellular $\text{H}^+$ & $K_\text{H}$ & $0.04565\ \si{fmol^{-1}}$ \\ 
		mem & Membrane capacitance & $C_m$ & $153400\ \si{fF}$ \\
		zF{\_}5 & Charge translocated by R5 & $z_5$ & $-0.0550$ \\
		zF{\_}8 & Charge translocated by R8 & $z_8$ & $-0.9450$ \\ \bottomrule& & 
	\end{tabular}
	\label{tab:bg_parameters}
\end{table}

\newpage
\section{Bond graph model structure}
\label{sec:BG_structure}
\begin{figure}[H]
	\centering
	\includegraphics[width=\linewidth]{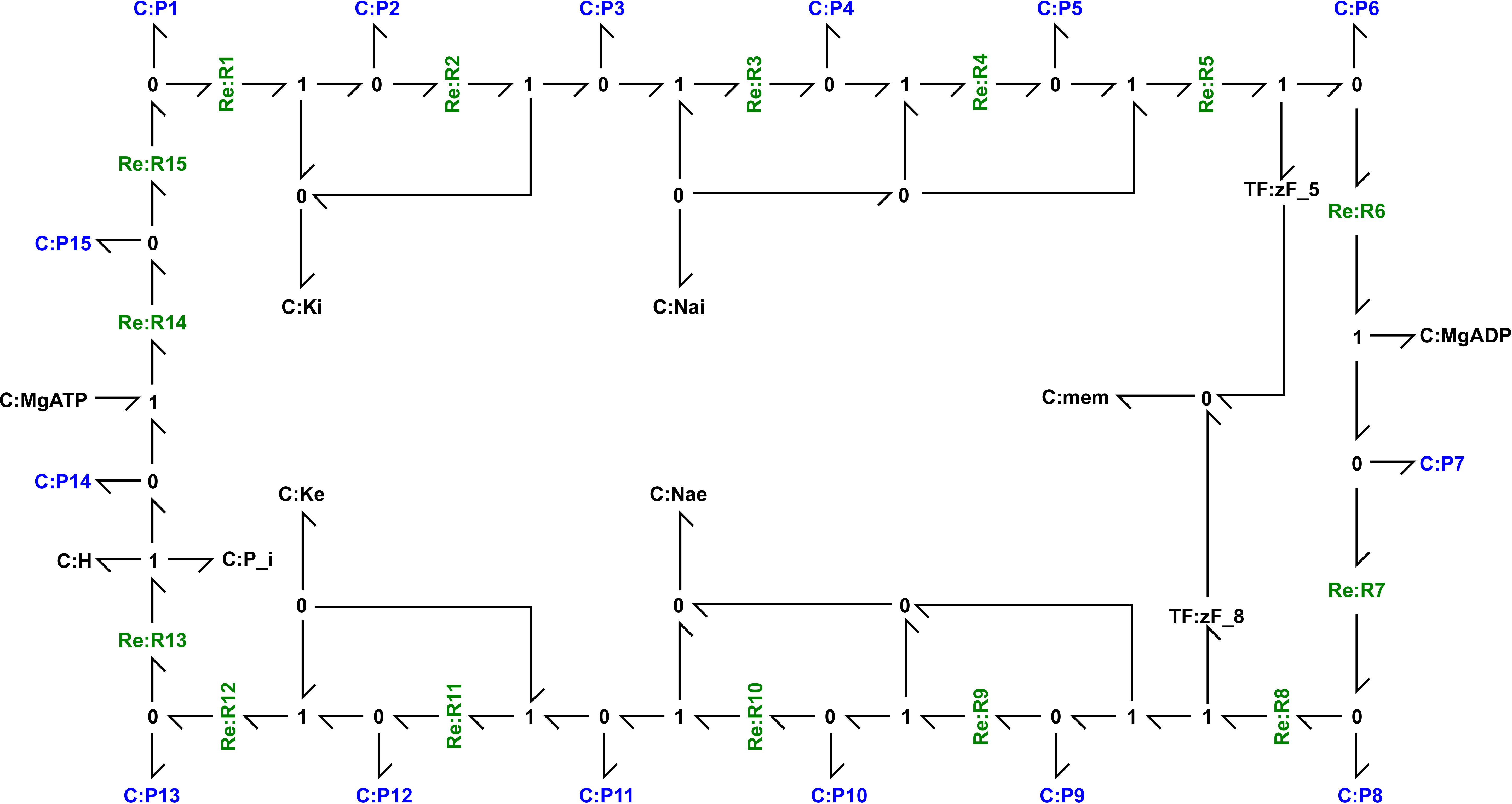}
	\caption{\textbf{Bond graph structure of the cardiac Na$^+$/K$^+$ ATPase model.} Pump states are coloured in blue, and reactions are coloured in green. The names for these components are consistent with their labels in \autoref{fig:NaK_scheme}.}
	\label{fig:Terkildsen_NaK}
\end{figure}

\end{document}